\documentstyle[aps,prl,floats,epsf,twocolumn]{revtex}

\begin{document}
\draft

\twocolumn[\hsize\textwidth\columnwidth\hsize\csname
  @twocolumnfalse\endcsname
  
\title{Persistent Currents and Dissipation
  in Narrow Bilayer Quantum Hall Bars}

\author{Jordan Kyriakidis and Leo Radzihovsky}
  
\address{Department of Physics, University of Colorado, Boulder,
    Colorado 80309}

\date{\today} 

\maketitle

\begin{abstract}
  Bilayer quantum Hall states support a flow of nearly dissipationless
  staggered current which can only decay through collective channels.
  We study the dominant finite-temperature dissipation mechanism which
  in narrow bars is driven by thermal nucleation of pseudospin
  solitons.  We find the finite-temperature resistivity, predict the
  resulting staggered current-voltage characteristics, and calculate
  the associated zero-temperature critical staggered current and gate
  voltage.
\end{abstract}
]

\narrowtext

A 2D electron gas bilayer, subjected to a strong perpendicular
magnetic field, can exhibit incompressible quantum Hall (QH) states
even for filling fractions corresponding to compressible states of
noninteracting layers~\cite{early_experiments}.  The nontrivial,
strongly interacting nature of these QH states lies in the fact that
they survive the limit of vanishing interlayer
tunneling~\cite{early_theory}.  They are stabilized by the exchange
part of the Coulomb interaction, which, in the limit of vanishing
single-particle tunneling, sets the scale of the gap and leads to
macroscopic interlayer phase coherence.

In addition to exhibiting the QHE for a uniform current, these states
support persistent currents that are counter-propagating in the two
layers with $J = J_{\text{top}} - J_{\text{bottom}}$.  In this Letter,
we study a thermally-driven decay mechanism of $J$ which controls the
current-voltage characteristics for staggered currents smaller than
the critical current $J_c$.  Because the bilayer system displays a
quantum Hall gap $\Delta$ in the phase-coherent ground state,
dissipation via single-particle mechanisms is strongly suppressed for
$k_B T \ll \Delta$.  Therefore, as with supercurrents in
superconductors~\cite{Langer}, the staggered-current decay rate is
dominated, for a range of parameters, by the collective mechanism of
soliton nucleation.

A convenient language for describing this strongly correlated
quantum-coherent gapped state is in terms of a pseudospin unit vector
field $\hat{m}(\vec{r}) = (\vec{m}_{\perp}, m_z)$~\cite{early_theory},
with $m_z=\cos\theta=n_{\text{top}}-n_{\text{bottom}}$ giving the
electron charge-density difference between top and bottom layers and
$\vec{m}_\perp=\sin\theta(\cos\phi,\sin\phi)$ characterizing the
relative phase $\phi=\phi_{\text{top}}-\phi_{\text{bottom}}$ of
electrons in two layers.  The energy functional describing long length
scale (larger than the magnetic length $\ell=\sqrt{\hbar c/B e}$)
variations of $\hat{m}(\vec{r})$ is given by~\cite{early_theory}
\begin{equation}
  \label{eq:1}
  H=\int d^2r\left[{\rho_s^\perp\over2}|\bbox{\nabla}\vec{m}_\perp|^2+
    {\rho_s^z\over2}|\bbox{\nabla}m_z|^2
    +\beta m_z^2 - \vec{h}\cdot\hat{m}\right]\nonumber,
\end{equation}
where electron Coulomb interaction is the origin of the effective
exchange constants $\rho_s^{z,\perp}$ that drive the transition into
the pseudo-ferromagnetic ground state, corresponding to the interlayer
phase coherent QH state.  The electrostatic capacitive energy $\beta$
introduces a hard-axis anisotropy, which forces the
pseudo-magnetization to lie in the $\perp$ plane ($m_z=0$) and thereby
reduces the full SU(2) pseudospin symmetry to
U(1)~\cite{early_theory,comment_lr}.  A combination of the external
gate voltage $V_g$ and the single-electron interlayer tunneling $t$
acts as an external pseudo-magnetic field
$\vec{h}=(t/2\pi\ell^2)\hat{x} + V_g\hat{z}$.  Because the tunneling
$t$ can be tuned independently of $\beta$ to be quite small, the
low-energy physics of this anisotropic QH pseudo-ferromagnet,
described by the Goldstone mode $\hat{m}$, can be fully explored
experimentally.

\begin{figure}
  \begin{center}
    \epsfxsize=8.5cm \epsffile{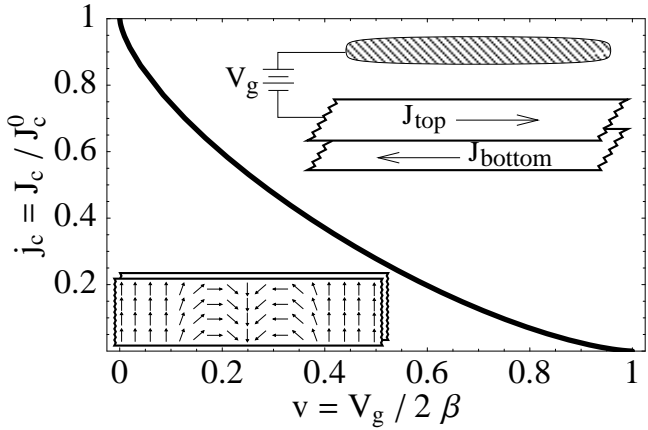}
  \end{center}
  \caption{Main Plot: Critical staggered current $j_c$ as a function
    of gate voltage $v$.  Upper Inset: Sample geometry considered in
    this work.  Lower Inset: Pseudospin configuration of the
    $2\pi$-line soliton whose nucleation leads to staggered current
    decay.}
    \label{fig:Jc}
\end{figure}

In the limit of vanishing tunneling $t$, an essentially exact
analytical treatment of narrow (1D limit) QH bars is possible and
leads to the following results.  The bilayer QH phase exhibits
staggered current-carrying states that are metastable and therefore
supports staggered persistent currents for $J<J_c(V_g)$, where the
critical current density is given by
\begin{mathletters}
\label{eq:Jc}
\begin{eqnarray}
  \label{eq:24}
  J_c(V_g) &=& J_c^0 q_*(v)
  \left[ 1 - v^2 /\big(1 - q_*^2(v)\big)^2 \right] \\
  &\stackrel{v \rightarrow 0}{\longrightarrow}&
  J_c^0 \left(1 - \left(v/4\right)^{2/3} \right),
\end{eqnarray}
\end{mathletters}
where $J_c^0 = 2\sqrt{2\beta\rho_s^\perp}/\hbar$ is the critical
current at zero gate voltage, which vanishes in the SU(2) invariant
$\beta\rightarrow0$ limit \cite{Ho}, $v=V_g/2\beta$ is a reduced
measure of the gate voltage $V_g$, and $q_*^2(v) = 1-v^{2/3}
[(\sqrt{4+v^2} + 2)^{1/3} - (\sqrt{4+v^2} - 2)^{1/3}]$.  At finite
temperature, $J_c(V_g)$, plotted in Fig.~\ref{fig:Jc}, therefore
delineates a low resistivity regime, where Ohmic dissipation is
dominated by slow thermal soliton nucleation, from a highly resistive
state dominated by quasi-particle dissipation.

\begin{figure}
   \begin{center}
     \epsfxsize=8.5cm \epsffile{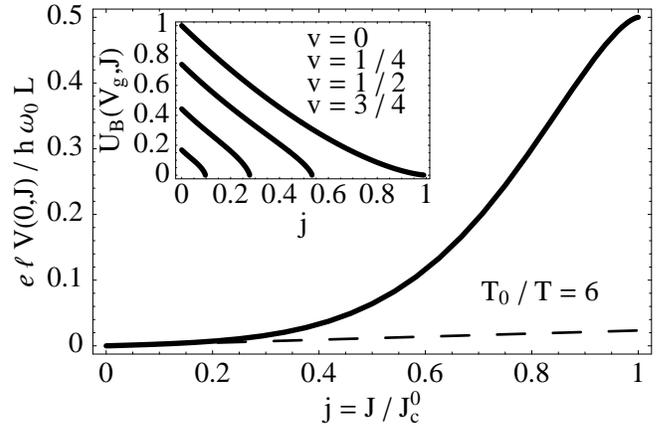}
   \end{center}
   \caption{Main Plot:  IV curve for $J<J_c(V_g)$, 
     $\rho_s^z=\rho_s^\perp$, $V_g=0$ and $T_0/T=6$.  Inset: Energy
     barrier for various gate voltages $v = V_g / 2 \beta$, with the
     barrier vanishing for $v \rightarrow 1$.  Qualitatively similar
     curves are obtained for the full range $0 \le
     \rho_s^z/\rho_s^\perp \le 1$.}
   \label{fig:iv}
\end{figure}

For $J<J_c(V_g)$, the staggered IV characteristics are plotted in
Fig.~\ref{fig:iv} and given by
\begin{equation}
  \label{eq:25}
  V(V_g,J) = \frac{h\omega_0 L}{e\ell}
  \sinh\left(\pi j T_0/2T\right)
  \text{e}^{-(U_B+\pi j / 2)T_0 / T},
\end{equation}
where $\omega_0 = \sqrt{2\pi\ell^2 t\beta} / \hbar$ is the microscopic
attempt frequency, $j=J/J_c^0$ is the reduced (dimensionless) current
density, $k_B T_0 = \hbar J_c^0 L_y$, with $L_y$ the narrow sample
dimension $\ll L_x \equiv L$, and $ E_B = k_B T_0 U_B$ is the
saddle-point energy barrier separating two different current-carrying
states.  The barrier is plotted in the inset of Fig.~\ref{fig:iv} and
is explicitly given by
\begin{eqnarray}
  \label{eq:barrier}
  U_B(V_g,J) &=&  \int_{m_1}^{m_0} \!\! dm
  \left\{
    \left(1 + \frac{\rho_s^z}{\rho_s^\perp} \frac{m^2}{1-m^2}\right)
  \right. \nonumber \\ &&
  \left. \mbox{} \times
    \left[ \left(m_0^2 - m^2\right)
      \left(1 - \left(\frac{qm_0}{m}\right)^2\right)
    \right.
  \right. \nonumber \\ &&
  \left.
    \left. \mbox{}
      + 2v \left(\sqrt{1-m_0^2} - \sqrt{1 - m^2}\right)
    \right]
  \right\}^{1/2},
\end{eqnarray}
where the limits of integration are $m_0^2 = j/q$ and $m_1^2 = q^2
[m_0^2 + 2 \bar{m}^2 (1 + \sqrt{1 + m_0^2 / \bar{m}^2})]$, with
$\bar{m}^2 = (1-m_0^2)(1-q^2)$, and the dimensionless wavevector $q$
is defined implicitly through the current $j = q [1 - v^2 / (1 -
q^2)^2]$.  The analytic expression for the barrier simplifies
considerably when one of its arguments vanishes.  We find that $U_B
(V_g,J)$ depends weakly on $\varepsilon \equiv 1 -
(\rho_s^z/\rho_s^\perp)$ and for $\varepsilon \rightarrow 1$ is given
by
\begin{mathletters}
  \begin{eqnarray}
    U_B (0,J) &=& {\pi\over4} (1-j)^2,\label{eq:32}\\
    U_B (V_g,0) &=& 
{1\over2}\left[\sin^{-1} \left(\sqrt{1-v^2}\right)-v \sqrt{1-v^2}\right],
 \label{eq:33}
 \end{eqnarray}
\end{mathletters}
and $ U_B(0,0) = \frac{1}{2} \sqrt{1-\varepsilon} -
\frac{1}{4\sqrt{\varepsilon}} \left( \frac{\pi}{2} -
  \sin^{-1}(1-2\varepsilon)\right)$.
For such narrow Hall bars the staggered linear resistivity $\rho$ is
always finite at finite temperature and is given by
\begin{equation}
  \label{eq:29}
  \rho(V_g) = \frac{V}{LJe} \stackrel{J\rightarrow 0}{\longrightarrow}
  \frac{h^2 \omega_0 L_y}{2 k_B T e^2 \ell}\,
  \text{e}^{-U_B(V_g,0)T_0/T}.
\end{equation}
From exact diagonalization studies~\cite{early_theory} at $d/\ell =
1/2$, we have $\beta = 5.1 \times 10^{-3} e^2/\ell^3\epsilon$ and
$\rho_s^\perp = 1.5 \times 10^{-2} e^2/\ell\epsilon$.  Taking in
addition $\ell=20\,\text{nm},\ T=1\,\text{K},\ \epsilon=13.2,\ 
t=0.1\,\text{meV},\ V_g=0,\ \text{and}\ \varepsilon=0$, we obtain
\begin{equation}
  \label{eq:resist}
  \rho(0) \approx 10^5 N (0.2)^N\, \Omega,
\end{equation}
where $N=L_y/\ell$.  Setting $N=5,\, 10,\, 15$ respectively gives
$\rho(0)=245\,\Omega,\, 0.2\,\Omega,\ \text{and}\ 115\,\mu\Omega$.

For a realistic system, there is a limited range of validity of the
above results, with other effects dominating outside of this range.
The constraint of quasi-equilibrium, which implies low decay rate,
together with the requirement that the thermal collective dissipation
mechanism dominates over single-particle current decay requires $k_B T
< E_B < \Delta$. At the same time, however, $T$ must be sufficiently
high so that thermal nucleation dominates over quantum tunneling of
phase slips.\cite{KRunpublished} Furthermore, in order for the bulk
nucleation rate to be experimentally observable, it is necessary that
it dominates over phase slips nucleated at surfaces, contacts, and
sample inhomogeneities.  Since the bulk nucleation rate scales with
the Hall bar length $L$, we expect that the bulk mechanism dominates
over surface nucleation for $L \gg \ell$.  Also, for the staggered
current decay rate to be dominated by the 1D {\em line} solitons
studied here, the saddle-point energy barrier given in
Eq.~(\ref{eq:barrier}) must be lower than barriers for the competing
mechanism of $\pm$ vortex pair nucleation.  For a sufficiently wide
Hall bar, the latter scenario will dominate, with the crossover
occurring for $L_y \approx \xi = \sqrt{\rho_s^\perp/\beta}$.

We now present the highlights of calculations that lead to these
results. Although quite distinct in detail, the spirit of our analysis
follows the classic work of Langer and Ambegaokar~\cite{Langer}.

The Euler-Lagrange (EL) equations for Eq.~(\ref{eq:1}) admit the
uniform current-carrying solutions
\begin{mathletters}
  \label{eq:uniform}
  \begin{eqnarray}
    \phi(k)&=&kx,\\
    \label{eq:2}
    m_\perp^2(v,q)&=&1 - v^2 / (1-q^2)^2,
    \label{eq:3}
  \end{eqnarray}
\end{mathletters}
where $q^2 = k^2 \rho_s^\perp / 2 \beta$ is the dimensionless wave
vector and we have taken our sample to lie in the $x$-$y$ plane with
dimensions $L_x\equiv L \gg L_y$.  Equation~(\ref{eq:3}) is valid in
the region $q^2 \le 1-v$ and $v = V_g / 2 \beta \le 1$.  (We consider
$V_g$ to be non-negative.)  The staggered current for this solution is
$J = 2 \rho_s^\perp m_\perp^2 k / \hbar$, or, equivalently, $j =
m_\perp^2 q$.

\begin{figure}
   \begin{center}
     \epsfxsize=8.5cm \epsffile{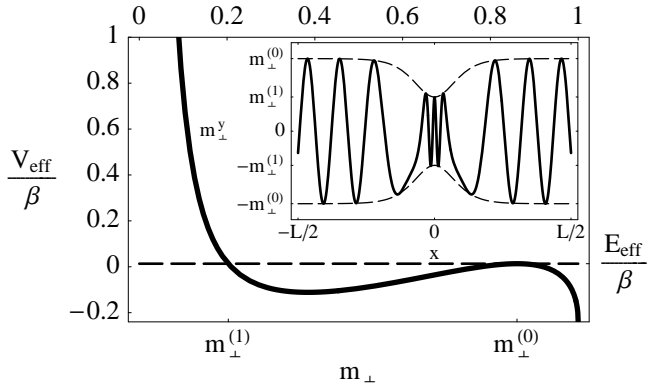}
   \end{center}
   \caption{Main Plot: Effective potential, Eq.~(\ref{eq:13}), for the
     mechanical analogy of the EL equations plotted for $v=0.5$ and
     $j=0.1$.  The physically stable configuration is the mechanically
     unstable point $m_\perp^{(0)}$.  Inset: Schematic drawing of the
     saddle-point (bounce) solution---the dominant contribution to the
     transition probability.}
   \label{fig:veff}
\end{figure}

For nonuniform solutions, the EL equations can be combined into a
single equation which, after some manipulation, can be written as
\begin{equation}
  \label{eq:12}
  \frac{1}{2} M \left(\partial_x m_\perp\right)^2 +
  V_{\text{eff}} = E_{\text{eff}},
\end{equation}
for some constant $E_{\text{eff}}$, with
\begin{mathletters}
  \label{eq:saddle}
  \begin{eqnarray}
    \label{eq:19}
    M\left(m_\perp\right) &=& \rho_s^\perp +
    \rho_s^z m_\perp^2 / (1-m_\perp^2), \\
    \label{eq:13}
    V_{\text{eff}}\left(m_\perp\right) &=& \beta
    \left( \frac{j^2}{m_\perp^2} -
      \left(1-m_\perp^2\right) +
      2v\sqrt{1-m_\perp^2}
    \right).
  \end{eqnarray}
\end{mathletters}
In the usual mechanical analogy, Eq.~(\ref{eq:12}) represents the
energy $E_{\text{eff}}$ of a particle at ``position'' $m_\perp$ and
``time'' $x$, moving in a potential $V_{\text{eff}}(m_\perp)$ and with
a space-dependent mass $M(m_\perp)$.  This potential is plotted in
Fig.~\ref{fig:veff} for $v=0.5$ ($j_c \approx 0.28$) and $j=0.1$.

The ``conservation of energy,'' Eq.~(\ref{eq:12}), immediately implies
the existence of two extended solutions \cite{comment}.  A uniform
current-carrying solution is given by Eq.~(\ref{eq:uniform}),
corresponding to the particle forever remaining on top of the hill at
$m_\perp^{(0)}$ with angular velocity $k$.  For this solution,
$\vec{m}_\perp$ winds azimuthally at a constant rate as a function of
$x$ with a constant amplitude $m_\perp^{(0)}$ deviating from the
equatorial plane ($m_\perp = 1$) with increasing $V_g$.  For fixed
winding $k$, this uniform current-carrying solution is a local minimum
of $H$ and is hence metastable.  Since the energy is lowered with
decreasing $k$, we expect at finite temperature the system will
thermally activate down to the $k=0$ zero-current ground state via
successive thermal transitions $k\rightarrow k-2\pi/L$.  To calculate
the rate of such transitions we need to compute the saddle-point
barrier separating two ``neighboring'' current-carrying metastable
states.  Brief reflection on the mechanical analogy shows that the
second solution is in fact a saddle-point solution corresponding to
the particle starting at ``position'' $m_\perp^{(0)}$ and ``time''
$x=-\infty$, spiraling down hill to $m_\perp^{(1)}$ at $x=0$ while
conserving angular momentum but increasing its angular velocity
$k(m_\perp)\propto1/m_\perp^2$, and finally bouncing back out to
$m_\perp^{(0)}$ for $x=+\infty$.  The resulting ``energy
conservation'' equation allows an exact determination of the
saddle-point solution, written in terms of a 1D integral,
\begin{equation}
  \label{eq:14}
  x = \frac{1}{\sqrt{2}} \int_{m_\perp^{(1)}}^{m_\perp(x)} \!\!
  dm_\perp \, \sqrt{ M(m_\perp) / \left[ E_{\text{eff}} -
    V_{\text{eff}}(m_\perp)\right]},
\end{equation}
for an appropriate value of $E_{\text{eff}}$, and is shown in the
inset of Fig.~\ref{fig:veff}.

The saddle-point solution is the nucleation site for the eventual
singularity where $m_\perp \rightarrow 0$ ($m^z\rightarrow \pm 1$),
and whereby the system can slip a loop reducing the total phase
winding $\Delta\phi$ by $2\pi$, and therefore leading to
staggered-current decay.  Contrary to what is often tacitly assumed,
the barrier for defect nucleation is {\em not} determined by the
energy of the defect---which may be either singular, as it is in
superconductors, or non-singular, as it is in the present case of the
full SU(2) space---although we expect them to be close in energy.
Instead, the barrier is controlled by a {\em non}-singular
saddle-point field configuration (inset of Fig.\ \ref{fig:veff}), which
is in the same topological sector as the current-carrying metastable
minimum.  It is also important to note that for sufficiently small
$\beta$, the energy for such phase slips is much smaller than the
quantum Hall gap $\Delta$ and that therefore the entire system remains
in the fully gapped quantum Hall state throughout the phase-slip
process.  This is to be contrasted with phase slips inside a
superconductor, where the order parameter, and therefore the bulk gap,
are suppressed within the vortex core and one in principle has to take
into account the core-confined low energy (normal) quasi-particle
degrees of freedom.

The energy barrier is defined as the difference in energy between the
saddle-point solution, Eq.~(\ref{eq:14}) and the uniform
current-carrying solution, Eq.~(\ref{eq:uniform}).  By exploiting the
mechanical analogy, and in particular the conservation law,
Eq.~(\ref{eq:12}), we obtain the expression quoted in
Eq.~(\ref{eq:barrier}).

In the steady-state regime, the staggered voltage $V(V_g,J)$ is
proportional to the net rate of phase slips, $\partial_t \Delta\phi /
2 \pi$, which itself is the difference between the rate of
current-decreasing transitions ($=\omega_0\exp[-E_B/k_BT]$) and the
rate of current-increasing transitions ($=\omega_0\exp[-(E_B+
\pi\hbar JL_y)/k_BT]$).  In a sample of length $L$, there are
approximately $L/\ell$ possible nucleation sites.  These
considerations lead directly to the expression in Eq.~(\ref{eq:25}).

In contrast to the energy barrier $E_B$, which is a static quantity,
we must consider pseudospin dynamics in order to compute the attempt
frequency $\omega_0$ appearing in Eq.~(\ref{eq:25}).  Within the
microscopic dynamical model valid at low temperatures, $T \ll t$, for
finite interlayer tunneling, $\omega_0^2$ is given by the ratio of the
curvature of the metastable well ($\sim t/2\pi\ell^2$) and the
dynamical mass term ($\sim \hbar^2 / 4 \pi^2 \ell^4 \beta$), leading
to $\hbar\omega_0=\sqrt{2\pi\ell^2t\beta}$. In contrast, for $T\gg t$,
we expect classical Langevin dynamics\cite{Langer}, characterized by a
kinetic ``drag'' coefficient $\gamma$, and which in the simplest
estimate gives $\omega_0=\beta/\gamma$.

When the transverse (narrow) dimension $L_y$ of the sample becomes
sufficiently large, the energy barrier $E_B \approx (\rho_s^\perp
\beta)^{1/2}L_y$ for nucleating a line-soliton defect becomes
comparable to the energy of nucleating a $\pm$ vortex pair---a
competing mechanism for inducing phase-slips.  Up to weak logarithmic
corrections, the energy of such a vortex pair is $E_v \approx
\rho_s^\perp = \beta \xi^2$, where $\xi = (\rho_s^\perp /
\beta)^{1/2}$ is the core size.  Vortex nucleation should therefore
dominate the 1D soliton nucleation considered here for $L_y > \xi$,
which can be tuned independently of the QH gap $\Delta$.  This is in
contrast with superconductors, where the corresponding 1D to 2D
crossover scale is the Ginzburg-Landau correlation length, controlled
by the superconducting gap.

Although it is difficult to extend our exact 1D analysis to the 2D
limit, we can estimate the staggered current decay rate using simple
scaling arguments. In 2D, the phase slip rate is controlled by $\pm$
vortex pair nucleation, analogously to superfluids and
superconductors. However, in contrast to those more familiar systems,
here vortices (half-skyrmion, i.e., merons) carry $\pm 1/2$
electromagnetic charge in addition to their $U(1)$ topological charge,
and there are therefore four elementary vortex defects $(+1/2, L)$,
$(-1/2, L)$, $(+1/2, R)$, and $(-1/2, R)$, with $L,R$ respectively
corresponding to $\pm 2\pi$ circulation of $\phi$. Correspondingly,
within the interlayer phase ($\phi$) coherent state, the $L$ and $R$
vortices are bound into $2$ types of topologically neutral pairs: (i)
electromagnetically neutral pairs $[(+1/2, L)$-$(-1/2, R)]$ and (ii)
electromagnetically charged pairs $[(+1/2, L)$-$(+1/2, R)]$
\cite{QHcomment}.  Nevertheless, we do not expect the Coulomb
interaction, which is subdominant to the topological-charge confining
potential, to play a role in staggered-current-induced vortex
ionization processes.  Hence, in the limit of vanishing $t$, standard
nucleation analysis for dissipation in superconducting
films\cite{Ambegaokar} can be easily extended to our system. It
predicts a highly {\em nonlinear} power-law staggered IV, $E\sim
J^\alpha$, with $\alpha(T)\geq3$ in the interlayer coherent state, a
result that contrasts strikingly with the {\em linear} staggered IV
found in the 1D limit.

Nonvanishing interlayer tunneling, $t$ explicitly breaks $U(1)$
symmetry and leads to nonuniform staggered current-carrying states
composed of a lattice of solitons of width $\delta = \ell
(2\pi\rho_\perp/t)^{1/2}$ and density $n$ akin to a periodic array of
Rayleigh-Benard current rolls.\cite{KRunpublished}.


In the dense soliton limit, $n\delta\gg 1$, (relevant for small $t$
and large $J$) the current $J(x)$ is nearly uniform and our 1D results
directly apply. In the dilute limit, $n\delta\gg 1$ (which is always
reached for sufficiently low
$J>J_{c1}\sim\sqrt{t}$\cite{AMunpublished,KRunpublished}), for
$\xi\ll\delta$, phase slips are confined to a single soliton, inside
which $\partial_x\phi\approx 2\pi/\delta$ is uniform and the tunneling
energy is on average zero. Hence, our 1D, $t=0$ analysis again applies
with the $J$-independent wavevector $k_{\text{eff}}=2\pi/\delta$.
Furthermore, scaling analysis\cite{KRunpublished} suggests that in the
opposite limit $\xi>\delta$, this solution is still valid, but with
the nucleation width set by $\delta$, rather than $\xi$.

In the 2D limit, the staggered current decays by ionization of $L$-$R$
vortex pairs, whose energy for $t>0$ grows {\em linearly} with the
separation $R$ as $(\rho_s t)^{1/2} R/\ell$. This therefore suggests
the existence of a true staggered {\em critical} current
$J_c=(e/\hbar)\sqrt{\rho_s t/\ell^2}$, with $E(J<J_c)=0$ even at
finite $T$ (up to corrections that vanish in the thermodynamic limit),
and for $J>J_c$, $E(J)\approx |J-J_c|^{\alpha(T)}$.


We thank Anton Andreev, Ramin Abolfath and Allan MacDonald for
discussions.  This work was supported by the NSF Grant \# DMR-9625111,
and by the A. P. Sloan and Packard Foundations.


\vspace{-0.5cm}

\begin{thebibliography}{10}
\vspace{-1.5cm}  
\bibitem{early_experiments} J.~P. Eisenstein {\it et~al.}, Phys.\ 
  Rev.\ Lett. {\bf 68}, 1383 (1992); S.~Q. Murphy {\it et~al.}, Phys.\ 
  Rev.\ Lett. {\bf 72}, 782 (1994).

\bibitem{early_theory} X.-G. Wen, A. Zee, Phys.\ Rev.\ Lett. {\bf 69},
  1811 (1992); K. Yang {\it et~al.}, 
Phys.\ Rev.\ B {\bf 54}, 11644 (1996); K. Moon {\it et~al.},
  Phys.\ Rev.\ B {\bf 51}, 5138 (1995).
  
\bibitem{Langer} J.~S. Langer, V. Ambegaokar, Phys.\ Rev. {\bf 164},
  498 (1967); D.~E. McCumber, B.~I. Halperin, Phys.\ Rev.\ B {\bf 1},
  1054 (1970).
  
\bibitem{comment_lr} Here, we have ignored the long-range part of the
  Coulomb interaction\cite{early_theory,AMunpublished}, which if kept
  forbids a detailed analytical treatment. The resulting short-ranged
  model is realized in bilayer systems with a screening conductor.

\bibitem{Ho} T. L. Ho, Phys.\ Rev.\ Lett. {\bf 73}, 874 (1994).
  
\bibitem{comment} Although an infinite number of solutions exist, only
  two satisfy the physically-motivated requirement that a saddle-point
  solution should be ``close'' in form to the uniform current carrying
  state, only deviating from it {\em locally}.
  
\bibitem{QHcomment} Interestingly, such an electromagnetically charged
  but topologically neutral vortex pair corresponds to a charged
  fermionic excitation which is gapped in the QH state. Although
  disorder will undoubtedly induce such charged quasi-particles, the
  QH state should survive as long as they remain localized. The
  delocalization of such topologically neutral vortex dipoles will
  destroy the QHE, but will preserve the interlayer phase-coherence,
  therefore suggesting the possibility of an exotic {\em gapless}
  interlayer phase-coherent state in a narrow sliver around $\nu=1/2$.
  L. R. thanks Steve Girvin for discussion on this point.
  
\bibitem{Ambegaokar}V. Ambegaokar, {\it et al.}, Phys.\ Rev.\ B {\bf
    21}, 1806 (1980).
  
  
\bibitem{KRunpublished} J. Kyriakidis and L. Radzihovsky, unpublished.
  
\bibitem{AMunpublished} R. Abolfath and A.~H. MacDonald, unpublished.

\end{thebibliography}
\end{document}